\documentclass[12pt]{article}
\usepackage{amsmath}
\usepackage{hyperref}
\usepackage{graphicx}
\usepackage{caption}
\usepackage{version}
\usepackage{float}
\usepackage{color}

\usepackage{tabularx}
\graphicspath{{./figures/}}
\begin{document}


\vspace*{.2cm}
\begin{center}
{\LARGE  {Role of specific growth rate in the development of different growth processes}}
\vspace*{1cm}
 \\ {\bf  {Dibyendu Biswas}\footnote{dbbesu@gmail.com} \bf{, Swarup Poria}\footnote{swarupporia@gmail.com} \bf{and Sankar Narayan Patra}\footnote{sankar.journal@gmail.com}}\\
{$^1$Department of Basic Science, Humanities and Social Science  (Physics)}\\
{ Calcutta Institute of Engineering and Management }\\
 {Kolkata-700040, India}\\
{$^2$Department of Applied Mathematics, University of Calcutta}\\
 {Kolkata-700009, India}\\
{$^3$ Department of Instrumentation Science, Jadavpur University}\\
 {Kolkata-700032, India}\\
\end{center}
\begin{center}
 \bf{Abstract}
\end{center}
~~~Effort has been given for the development of an analytical approach that helps to address several sigmoidal and non-sigmoidal growth processes found in  literature. In the proposed approach, the role of specific growth rate in different growth processes has been considered in an unified manner. It is found that the different growth equations can be derived from the same functional form of rate equation of specific growth rate. A common functional form of growth and growth velocity have been derived analytically and it has been shown that different values of the parameters involved in the description lead to different growth function. The theta logistic growth can be explained in this proposed framework. It is found that the competitive environment may increase the saturation level of population size. 
\section{Introduction}\paragraph{}
\label{sec:1}
Growth is  an extremely complex  nonlinear phenomenon observed in the field of biology, economics and other natural sciences. Several models {\cite{Hershey,Hart,Stearns,verhulst,may,Tsoularis,Richards,Gilpin1973,Bertalanffy1957}} have been developed in order to describe different growth processes. The exponential growth model, one of the simplest population growth model, does not consider the limitation of resources imposed by the environment. Such growth model is generally unrealistic to describe a biological system although it is very useful to describe  different physical systems, i.e; autocatalytic reaction, radioactive decay, bacterial cloning {\cite{Hershey}} etc. Potential growth, other than the exponential growth, also shows non-saturated growth and does not reach carrying capacity. Potential growth function is extensively used in different fields  {\cite{Hart,Stearns, Calders,Brown,Day}}.

Logistic growth model, in continuous {\cite{verhulst,pearl}} or discrete {\cite{may}} form, includes the initial exponential nature of growth rate and competition under limited resources described by saturation values. Discrete logistic growth equations are able to describe chaotic behaviours of the system {\cite{Hanski}}. The continuous form of logistic growth models do not show intrinsic bifurcations, and as a result, it is much more easier to handle analytically. The inflection point of such model is fixed and always takes place at the population size that is half of the saturation value. This imposes undesirable restriction on the shape of the characteristic curve. Though, it forms the basis of several extended models {\cite{Tsoularis,Turner,Blumberg}}. Logistic growth model is extensively used in the field of biology {\cite{Morgan,Pearl1930,Krebs}} and many other fields  {\cite{Fisher,Marchetti,Herman}}.

A slightly generalized version of the logistic growth model, termed as $\theta$-logistic growth model, is also introduced to study plant growth {\cite{{Richards,Nelder}}, population ecology {\cite{Gilpin1973,Gilpin1976}}, avian population dynamics {\cite{Saether(a)2002}}, environmental stochasticity on population growth {\cite{Saether(b)2002}}, species abundance in community ecology {\cite{Diserud}}. Similar to logistic growth model, the population in this type of growth model converges with time to the same saturation level. In this model, a new parameter ($\theta$) representing intra-specific competition regulates the time required to reach its saturation value, may be termed as carrying capacity. $\theta$ also indicates nature of fluctuation due to environmental stochasticity. Large fluctuations in population size are observed when $\theta$ is small, whereas more stable fluctuations are found in case of larger values of $\theta$ {\cite{seather}}.

The growth models showing saturated nature other than logistic and $\theta$- logistic growth model are von Bertalanffy model {\cite{Bertalanffy1957,Bertalanffy1966}}, Kleiber's law of growth {\cite{Kleiber}}, Gomperzian growth {\cite{Gompertz}} etc. von Bertalanffy model is frequently used in different types of allometric modelling. Kleiber's law of growth is recently successfully used by West $et$ $al.$ {\cite{West}} to derive a differential equation showing growth of biological masses. Gompertz growth model, formulated to model human demographic data, is also frequently used in modelling tumour growth {\cite{Norton}}.

Growth in its various forms have been studied for centuries. Although there are a large number of well-known growth laws, but they are almost entirely empirical and lacks a theoretical foundation. Some attempts have been made to address these models from some basic principles. These attempts include derivations based on the framework of synergistic system {\cite{Savageau1979,Savageau1980}}, biological hierarchies {\cite{Witten1985}}, cell kinetics {\cite{Gyllenberg1989}}, entropy-change {\cite{Calderon1991,Ling1993}}, cell population heterogenity and statistical mechanics {\cite{Kendal1985}}. In each case, at least one assumption is not biologically founded {\cite{Bajzer1997,Bajzer2000}}. Therefore, a  biologically convincing, simple and mathematically sound argument for different growth models is still required.

The analysis and comparison of different types of growth model in a unified manner is expected for better understanding of growth processes. Savageau {\cite{Savageau1979,Savageau1980}} tried to address this issue and derived a general growth equation by considering the properties of underlying determinants of a complex system. Tsoularis $et$ $al.$ {\cite{Tsoularis}} tried to address this issue by proposing generalized logistic growth model, but the proposed form of generalized growth model is empirical. They have shown that different types of growth models can be derived as a special case of their proposed functional form. Later, Thoularis-Wallace model is explained by Postnikov with the help of three-compartmental demographic model {\cite{Postnikov}}. Castorina $et$ $al.$  also tried to address Gompertz and West-type growth in terms of adimensional analysis {\cite{Castorina}}. They have failed to describe logistic growth {\cite{biswas_poria}}. Kozusko $et$ $al.$ presented unified approach for sigmoid tumour growth based on cell proliferation and quiescence {\cite{Kozusko2007}}. They have basically considered the behaviour a population originated from the interaction of two subpopulations and presented a set of sigmoidal growth curve {\cite{Kozusko2011}}. It is not possible to address non-sigmoidal growth curves with the help of this approach.  These facts motivated us to propose  a unified analytical approach based on change of specific growth rate with time. The proposed approach is able to capture different types of sigmoidal and non-sigmoidal growth curves.

In the present communication, different types of growth models including $\theta$-logistic growth model are addressed from the most fundamental growth equation. The fundamental growth equation must be valid for different types of growth processes. A common functional form of growth (or growth velocity/ rate), similar to Richard's growth function, has been derived based on the proposed approach. It has been shown that different types of growth function (or growth velocity) could be derived from the same functional form for different values of parameters involved in the description. It is also found that different growth processes are following a common functional form of rate of specific growth rate. It is also observed that a competitive environment may be responsible for the increase in population size. In this connection, the variation in growth features of different growth processes have been studied in terms of different parameters involved in this approach.

The paper is organized as follows: In Sec. II, we first propose a description of a growth process in a generalized way. We have derived generalized growth equation and generalized growth velocity. In Sec. III, we have reported the dependence of growth features of different growth processes on the parameters of the proposed description.  A possible explanation of the parameters involved in the description of growth processes is given in terms of reproduction processes and energy consumption (or a competitive environment that influences energy consumption) of a growing system. Finally we conclude about our result in section IV.
\section{The role of specific growth rate and generalized growth equation}\paragraph{}
\label{sec:2}
Growth is a natural process which may be addressed by two state variables, that are (i) the observed physical quantity ($x$) of interest, and (ii) the specific growth rate ($s$) of the physical quantity ($x$). The growth dynamics originates from the precise relationship between $x$ and $s$ {\cite{Sibly}}.   Therefore, the growth or evolution of any physical quantity ($x$) with time ($t$) in a physical process can be expressed in the following form,
\begin{equation}
\frac{dx}{dt}=s(x,t)x(t)
\label{eh:1}
\end{equation}
Where, $s(x,t)$ is the specific growth rate of the variable $x(t)$. $x(t)$ may vary with some other characteristic variable of the system. But we should consider here the evolution of a growing system with respect to time. Thus, an ordinary differential equation, instead of partial, would serve the purpose.

Both, $x$ and $s$, are function of time. In a growth process, $x$ is expected to increase with time. It is justified to consider that specific growth rate declines with increasing $x$ {\cite{Sibly}} or with time ($t$). Thus, it is expected that the rate of change of specific growth rate (i.e, $\dot{s}$) must be less than zero in a growth process. Therefore,
\begin{equation}
\frac{ds}{dt}< 0
\label{eh:2}
\end{equation}
Hence, it would be better to introduce a variable $R(s)$ that can be expressed as,
\begin{equation}
R(s)=-\frac{ds}{dt}.
\label{eh:3}
\end{equation}
We assume that $R(s)$ can be expressed in terms of power series as,
\begin{equation}
R(s)={\sum_0^\infty} p_n s^n
\label{eh:4}
\end{equation}
The polynomial in equation (\ref{eh:4}) could be truncated up to any terms and the truncated series could be used in calculation to represent any natural phenomenon. One or more than one terms in that truncated series may be equal to zero. It has been shown in the dimensionless analysis that a truncated series up to third terms can produce oscillatory growth (or decay). When the second terms of the same truncated series is zero, it can produce sustained oscillation {\cite{Biswas2016}}.

Here, we have considered the series up to third terms and assumed that the first term ($p_0$) of the truncated series is zero.  Therefore, the truncated series generates a quadratic form as a special case. It would be shown in the proceeding section that such a truncated series leads to a growth equation similar to Richard's growth function {\cite{Richards}}.

By integrating equation ({\ref{eh:3}}) for $R(s)=p_1s+p_2s^2$, specific growth rate ($s$) can be expressed as,
\begin{equation}
s=\frac{p_1s_0}{[(p_1+p_2s_0)\exp (p_1t)-p_2s_0]}
\label{eh1:4}
\end{equation}
where, $s_0$ is a constant defined as $s=s_0$ at $t=0$.

From equations ({\ref{eh:1}}) and ({\ref{eh:3}}), the following relation can be established,
\begin{equation}
\ln x=\int sdt+ C_1
\label{eh2:5}
\end{equation}
where, $C_1$ is a constant of integration.

Equation ({\ref{eh2:5}}) can be expressed for $R(s)=p_1s+p_2s^2$ as,
\begin{equation}
x^{p_2}=\frac{x_0^{p_2}(p_1+p_2s_0)}{p_1+p_2s}
\label{eh3:5}
\end{equation}
where, $x_0$ is a constant defined as $x=x_0$ at $t=0$.

Eliminating $s$ from equation ({\ref{eh3:5}) with the help of  equation ({\ref{eh1:4}}), the following relation can be established,
\begin{equation}
x=
\Big[\frac{x_0^2(p_1+p_2s_0)}{p_1}-\frac{x_0^2p_2s_0}{p_1\exp (p_1t)}\Big]^{\frac{1}{p_2}}
\label{eh:5}
\end{equation}
and, growth rate ($\frac{dx}{dt}$) can be expressed as
\begin{equation}
\frac{dx}{dt}=x_0s_0x^{1-p_2}\exp(-p_1t)
\label{e1:6}
\end{equation}
Eliminating $t$ from equation  ({\ref{e1:6}}) using equation ({\ref{eh:5}}), growth rate can  be expressed as,
\begin{equation}
\frac{dx}{dt}=\frac{x_0^2(p_1+p_2s_0)}{p_2}x^{1-p_2}-\frac{p_1}{p_2}x
\label{e:6}
\end{equation}
Equation ({\ref{eh:5}}) [or equation ({\ref{e:6}})] is derived analytically based on the concept of rate of change of specific growth rate. It may be treated as a generalized growth function because most of the well-known growth models could be derived (presented in the next section) from it for different values of $p_1$ and $p_2$.
\section{Result and discussion}\paragraph{}
\label{sec:3}
The approach proposed here is enriched with several quantitative solutions related to growth mechanisms found often in literature.  This analytical approach is useful to represent different types of sigmoidal and non-sigmoidal growth along with linear growth.

It is expected that the state variable ($x$) must reach a saturation level ($x_{max}$) for different types of sigmoidal growth for which $x_{max}>0$ and $s=0$.  To satisfy these conditions, the following condition must be obeyed by the growing system,
\begin{equation}
p_1>-p_2s_0
\end{equation}
Environmental changes and adoptive strategies are well-known to unbalance equillibrium of a population. As a result of such perturbations, a change in $p_1$ and/or $p_2$ of the system is expected. The system then would try to reset its saturated value (governed by the value of $p_1$ and $p_2$) to a new level and it would show growth (or decay) to attain that level. If the perturbation is within the tolerance limit, the system will follow a new growth dynamics based on the value of $p_1$ and $p_2$. Such deviation from saturated level and attainment of new saturation level could be explained by means of proposed unified approach for different growth mechanisms. The basis of such unified approach proposed here is change of specific growth rate with time. Therefore, it is possible to address attainment of different saturation levels by a growing system at different instant of time by considering a change in $p_1$ and $p_2$.

We should now consider the effect of variation of parameters related to different growth processes and a possible interpretation of those parameters based on the same functional form of rate equation of specific growth rate.
\subsection{Potential growth }\paragraph{}
From equation ({\ref{e:6}}), it can be shown that $\frac{dx}{dt}\rightarrow x_0^2s_0x^{1-p_2}$ when $p_1\rightarrow 0$. The physical quantity ($x$) can be expressed for this condition as,
\begin{equation}
x\rightarrow (x_0^{p_2}+x_0^2s_0p_2t)^{\frac{1}{p_2}}
\label{eh:18}
\end{equation}
It represents the potential growth function used in tumour biology {\cite{Hart}}, lifehistory theory {\cite{Stearns}}.
It leads to linear growth with time for the conditions: $p_1\rightarrow 0$ and $p_2=1$.

Potential growth is found to be observed in different physical system {\cite{Hart,Stearns,Brown}}. It could be considered as one of the limiting case of the proposed rate equation of specific growth rate represented by $\frac{ds}{dt}=-(p_1s+p_2s^2)$. According to this proposed description, it depends on only $p_2$. As a result, it can be concluded that only one type of growth mechanism is dominant in this type of growing system (it is considered in detail in next paragraph).  It is represented by first and third characteristic line, from the top, of figure \ref{figh:1}. The second characteristic line from the top of figure \ref{figh:1} represents linear growth ($p_2=1.0$). According to this proposition, it can be treated as a special case of potential growth.

The main issue of life-history theory is the allocation of energy consumed by an organism for different types of adaptive strategies in distinct stages of life. Reproduction and survival are important factors in determining growth of an organisim {\cite{Stearns}}. Consumed energy (governed by competitive environment) is mainly used for growth in non-reproductive stage of the organism. At this stage, the allocation of consumed energy depends on the size of the organism and follows a potential growth function {\cite{Stearns}}. The allocation of energy for reproduction is negligible ($p_1\rightarrow 0$). Less amount of energy is allocated for growth when reproduction gains profound importance. The consumed energy is then distributed between survival and reproduction. The value of $p_2$ may be related to the strategy of consumption of energy by the organism. Therefore, it may be concluded from the above consideration that the parameter $p_1$ may be related to reproduction processes whereas $p_2$ is related to energy consumption of an organism for growth (may be effected by competition). It is also found that the lower value of $p_2$ initiates the higher growth rate (as shown in figure \ref{figh:1}). Higher growth rate is initiated by higher value of consumed energy. Therefore $p_2$ could be treated as a measure of energy consumed by an organism and/or intra-specific competition. It is expected that higher degree of intra-specific competition would lower energy consumed by organism, that in turn lowers the energy allocated for growth. Therefore, $p_2$ could also be treated as a measure of intra-specific competition and inversely related to the energy consumed by the organism.
\subsection{ $\theta-$ logistic growth}\paragraph{}
When $p_2<0$, say $p_2=-\beta$, equation ({\ref{e:6}}) can be expressed as,
\begin{equation}
\frac{dx}{dt}=\frac{p_1}{\beta}x\Big[1-\frac{x^\beta}{\frac{x_0^\beta p_1}{p_1-\beta s_0}}\Big]
\label{eh:7}
\end{equation}
When $p_1-\beta s_0\neq 0$, equation ({\ref{eh:7}}) can be expressed as,
\begin{equation}
\frac{dx}{dt}=\frac{p_1}{\beta}x\Big[1-\frac{x^\beta}{K^\beta}\Big]
\label{eh:8}
\end{equation}
Where, $\frac{x_0^\beta p_1}{p_1-\beta s_0}=K^\beta$.
It is the form of $\theta$-logistic growth {\cite{Nelder}}. It shows more accurate results than usual logistic growth {\cite{Gilpin1973}}. It is also used to measure fluctuations originated from environmental stochasticity {\cite{seather}}.

In this proposed description of growth processes, $\theta-$ logistic growth could be addressed with the condition $p_2<0$.  The effect of variation of $p_2$ on growth rate is represented by figure \ref{figh:2}. All of the characteristic curves are logistic by nature. Among them, second from the top stands for usual logistic growth. It is found that the growth rate at the early stage, optimum growth rate, optimum size {\cite{Biswas}} and $x_{max}$ are effected by $p_2$. The value of $x_{max}$ decreases for the condition $-1\leq p_2 <0$. It increases for the condition $p_2<-1$. Increase in initial growth rate and optimum growth rate are observed with the increase in $p_2$. Such dependence on $p_2$ can be interpreted in the following way: the magnitude of $p_2$ can be treated as a measure of intra-specific competition. If $p_2<-1$ then the intra-specific competition is very high. As a result, the system takes more time to reach saturation level. It in turn lowers the optimum growth rate. If $-1<p_2<0$, then the competition is lower. Therefore, higher value of optimum growth rate and $x_{max}$ are expected. If $p_2=-1$ then the competition is moderate. It is also found in this study that higher value of $x_{max}$ may be observed even in case of higher intra-specific competition than that for a moderate competition. It is not normally expected. This is one of the important findings of this proposed analysis. This finding may be helpful for the researchers for better understanding of the effect of competitive environment on growth mechanism.
\subsubsection{Logistic  growth}
Setting  $\beta=1$ or $p_2=-1$ in equation ({\ref{eh:8}}), the usual logistic growth can be obtained as,
\begin{equation}
\frac{dx}{dt}=p_1x\Big[1-\frac{x}{K}\Big]
\label{eh:8a}
\end{equation}
Figure \ref{figh:3} shows the variation in growth rate with state variable $x$ for different values of $p_1$ in case of usual logistic growth ($p_2=-1$). It shows that growth rate does not depends on $p_1$ at the initial stage of growth. But the rate of change of growth rate with respect to state variable $x$ changes with $p_1$. It increases with the decrease in $p_1$. As a result, the saturated value ($x_{max}$) of the state variable $x$ (for which $\frac{dx}{dt}=0$) decreases with the increase in $p_1$. It is also found that the optimum growth rate and the optimum size {\cite{Biswas}} increase with the decrease in $p_1$.
\subsubsection{Gompertzian  growth}\paragraph{}
Equation ({\ref{eh:8}}) can be expressed as,
\begin{equation}
\frac{dx}{dt}=\frac{p_1}{\beta}x\Big[1-\exp{\Big(\beta \ln\frac{x}{K}}\Big)\Big]
\label{eh:11}
\end{equation}
\begin{equation}
=p_1x\Big[-\ln{\frac{x}{K}}-\beta\Big(\ln\frac{x}{K}\Big)^2\Big]
\label{eh:12}
\end{equation}
When $\beta\rightarrow 0$, equation ({\ref{eh:12}}) can be expressed as,
\begin{equation}
\frac{dx}{dt}=p_1x\ln{\frac{K}{x}}
\label{eh:13}
\end{equation}
Equation ({\ref{eh:13}}) represents Gompertz law of growth. It was first introduced to evaluate mortality table {\cite{Gompertz}}. It is also frequently used for the description of tumour growth {\cite{steel,wheldon}}.

Figure \ref{figh:4} represents different types of Gompertz type growth described by $\frac{ds}{dt}=-(p_1s+p_2s^2)$ with $p_2\rightarrow 0$. It shows dependence of growth feature on $p_1$. It is found that the optimum growth rate {\cite{Biswas}} increases with the decrease in $p_1$. But it does not alter the optimum size and $x_{max}$.

Different types of mathematical computation models are used to study growth features of tumours {\cite{wheldon}}. In vitro and experimental studies show that the growth of tumour follows Gompertz law of growth {\cite{Norton}} and attains a saturation level. The Gompertz law of growth ($p_2\rightarrow 0$) indicates that the consumed energy in tumour is entirely used for reproduction ($p_1\neq 0$). The condition $p_2\rightarrow 0$ indicates that the intra-specific competition is negligible. In other word, the degree of cooperation, a measure of self-organization, is very high. It is in accordance with the research work of Molski $et$ $al.$ {\cite{molski}}. The self-organization is also related to the coherent state of the system. such coherence is confirmed by Gomtertzian regression rate of tumour {\cite{molski}} that is found in case of external perturbations {\cite{valder}}. Therefore, it can be concluded that the parameter $p_1$ may be related to the reproduction process of the system.
\subsection{Von Bertalanffy and Kleiber's growth}\paragraph{}
When $0<p_2<1$ and $p_2=0.34$, equation ({\ref{e:6}}) can be expressed as,
\begin{equation}
\frac{dx}{dt}=\alpha_1 x^{0.66}-\alpha_2x
\label{eh:9}
\end{equation}
Where, $\alpha_1=\frac{x_0^2(p_1+0.34s_0)}{0.34}$ and $\alpha_2=\frac{p_1}{0.34}$.

This is known as von Bertalanffy growth equation introduced to model fish weight growth {\cite{Bertalanffy1938}}. It is based on a simple assumption that the energy consumed by an organism is proportional to the surface area of the body of that organism. It is basically a modification of Verhulst logistic growth and can be treated as a special case of Bernoulli's differential equation.

Again, equation ({\ref{e:6}}) can be expressed for $p_2=0.25$ as,
\begin{equation}
\frac{dx}{dt}=\alpha_1 x^{0.75}-\alpha_2x
\label{eh:10}
\end{equation}
Where, $\alpha_1=\frac{x_0^2(p_1+0.25s_0)}{0.25}$ and $\alpha_2=\frac{p_1}{0.25}$.

This is related to Keblier law of growth that is successfully used by West $et$ $al.$ to describe the evolution of biological masses {\cite{West}}. They have considered fractal-like distribution of resources in a living organism. The ontogenetic growth model proposed by West $et$ $al.$ can successfully describe the growth of any living organism, from protozoa to mammals.

Both, West-type and von Bertalanffy type growth, are used to describe growth of organisms. According to proposed framework, the condition $\frac{ds}{dt}=-(p_1s+p_2s^2)$  with $p_2=0.25$ represents West-type ontogenetic growth. The variation of $p_1$ for West-type growth is shown in figure \ref{figh:5}. It is found  that the optimum growth rate and $x_{max}$ decrease with the increase in $p_1$. The same is true in case of von Bertalanffy type growth that is described by $\frac{ds}{dt}=-(p_1s+p_2s^2)$  with $p_2=0.34$ (as shown in figure \ref{figh:6}). A graphical comparison between von Bertalanffy type growth and West type biological growth is considered in figure \ref{figh:6} for same $p_1$. It is found that the specific growth rate and the optimum growth rate for von Bertalanffy type growth are greater than that of West-type growth in case of same $p_1$. But the optimum mass and $x_{max}$ of von Bertalanffy are lower than that of West type growth for same $p_1$.

In case of West type or von Bertalanffy type growth, $p_1$ and $p_2$ are not equal to zero. The value of $p_2$ in case West type growth is lower than that of von Bertalanffy type growth. Therefore, it may be concluded that the ability of energy consumption is greater for West type growth than that for von Bertalanffy type growth. The lower value of growth rate for the same value of state variable ($x$) and $p_1$ for West type growth may be due to higher value of metabolic cost of survival. The nonzero value of $p_1$ in both cases may be an indicative of allocation of energy for reproduction from the beginning. The non-zero value of $p_1$ and $p_2$ indicates the coexistence of cooperation and competition mechanism along with reproduction in the cellular level of a growing organism.
\subsection{Monomolecular growth}\paragraph{}
Monomolecular growth is also termed as limited growth or growth of decreasing potential{\cite{Savageau1980}}. It is found in certain animals during most of their lives {\cite{Bertalanffy1960,Brody1964}} and certain plants {\cite{Richards1969}}.
When $p_2=1$, equation ({\ref{e:6}}) can be expressed as,
\begin{equation}
\frac{dx}{dt}=p_1\Big[\frac{x_0(p_1+s_0)}{p_1}-x\Big]
\label{eh:20}
\end{equation}
By considering $x_{max}=\frac{x_0(p_1+s_0)}{p_1}$, equation ({\ref{eh:20}})  can be expressed as
\begin{equation}
\frac{dx}{dt}=p_1[x_{max}-x]
\label{eh:21}
\end{equation}
Equation ({\ref{eh:21}}) is termed as Monomolecular growth.

In the proposed description, monomolecular growth is found for $p_2=1$. The maximum attainable value ($x_{max}$) of the physical quantity is governed by $p_1$ only. As $p_2$ may be treated as a measure of energy consumption or competition,  the energy consumption is expected to be less than West-type and  Bertalanffy-type growth. The degree of cooperation (may be related to $p_1$) may depend on organism. Therefore, the degree of cooperation dictates the value of $x_{max}$ in this case.

Conditional evolution of different growth equations from the proposed functional form of rate equation of specific growth rate is presented in table ($1$).\\
\begin{table}[h]
\caption{Conditional evolution of different growth functions from same rate equation of specific growth rate.}
\centering
\begin{tabular}{c c c c}
\hline\hline
Rate of specific & Initial values of   & Conditions & Nature of\\ [0.5ex]
growth rate & state variables &  & growth\\ [0.5ex]
\hline
& & $p_1>0, p_2=1, p_1+s_0p_2> 0$ &Monomolecular\\
& & $p_1>0, p_2\rightarrow 0, p_1+s_0p_2> 0$ &Gompertzian\\
& & $p_1\rightarrow 0, p_2\neq 0$ &Potential\\
$p_1s+p_2s^2$& $s_0>0$& $p_1>0, p_2=-1, p_1+s_0p_2> 0$ &Logistic\\
& $x_0>0$ & $p_1>0, p_2<0, p_1+s_0p_2> 0$ &$\theta$-logistic\\
& & $p_1>0, p_2=0.25, p_1+s_0p_2> 0$ & West-type, Keblier \\
& & $p_1>0, p_2=0.34, p_1+s_0p_2> 0$ &von Bertalanffy\\
& & $p_1\rightarrow 0, p_2=1.0$ &Linear\\
\hline
\end{tabular}
\label{t:6}
\end{table}
The differential equation (\ref{e:6}) is very similar to Richard's growth function. It is not empirically presented in this communication. To derive this expression, we have presented here an analytical approach based on the concept of change of specific growth rate. The approach can explain the variation of parameters and scaling coefficients of Richards growth function in terms of two coefficients ($p_1$ and $p_2$) related to the change of specific growth rate. It also shows a possible explanation of $p_1$ and $p_2$ . It correlates maximum attainable value of the state variable ($x_{max}$) with its specific growth related coefficients. Therefore, it can be concluded that this analytical approach should be helpful to compare different growth processes in the lime light of Richard's growth function. It also show that competitive environment may enhance the saturation value of state variable. In brief, these are the merits of this communication.
\section{Conclusions}\paragraph{}
\label{sec:5}
The purpose of this communication is to present a generalized approach that could accommodate several well-known growth models. We have identified that several types of growth models could be derived form the same functional form of rate equation of specific growth rate for different conditions. A common functional form of growth function (or growth rate) has been derived analytically. Potential, Gompertzian, monomolecular, West-type and von Bertalanffy, logistic and $\theta$-logistic  growth functions (or growth rates) could be obtained from the proposed form. It is shown that this analytical approach is also able to address linear growth. The effect of variations of the different relevant parameters on growth rate and size of a physical quantity of interest have been studied in detail. It is also observed that the intra-specific competition may enhance the saturated value of the state variable in some cases. The key observations of this communication are: (i) the increase of  population size in a competitive environment; and (ii) analytical approach to  describe $\theta$-logistic growth  and other sigmoidal and non-sigmoidal growth functions along with possible explanations of significant observations regarding growth processes. Our study may be useful for understanding the basic mechanism behind  different biological and social growth processes.\\

\newpage
\begin{figure}
  \centering
  \includegraphics[width=4in, height=2in]{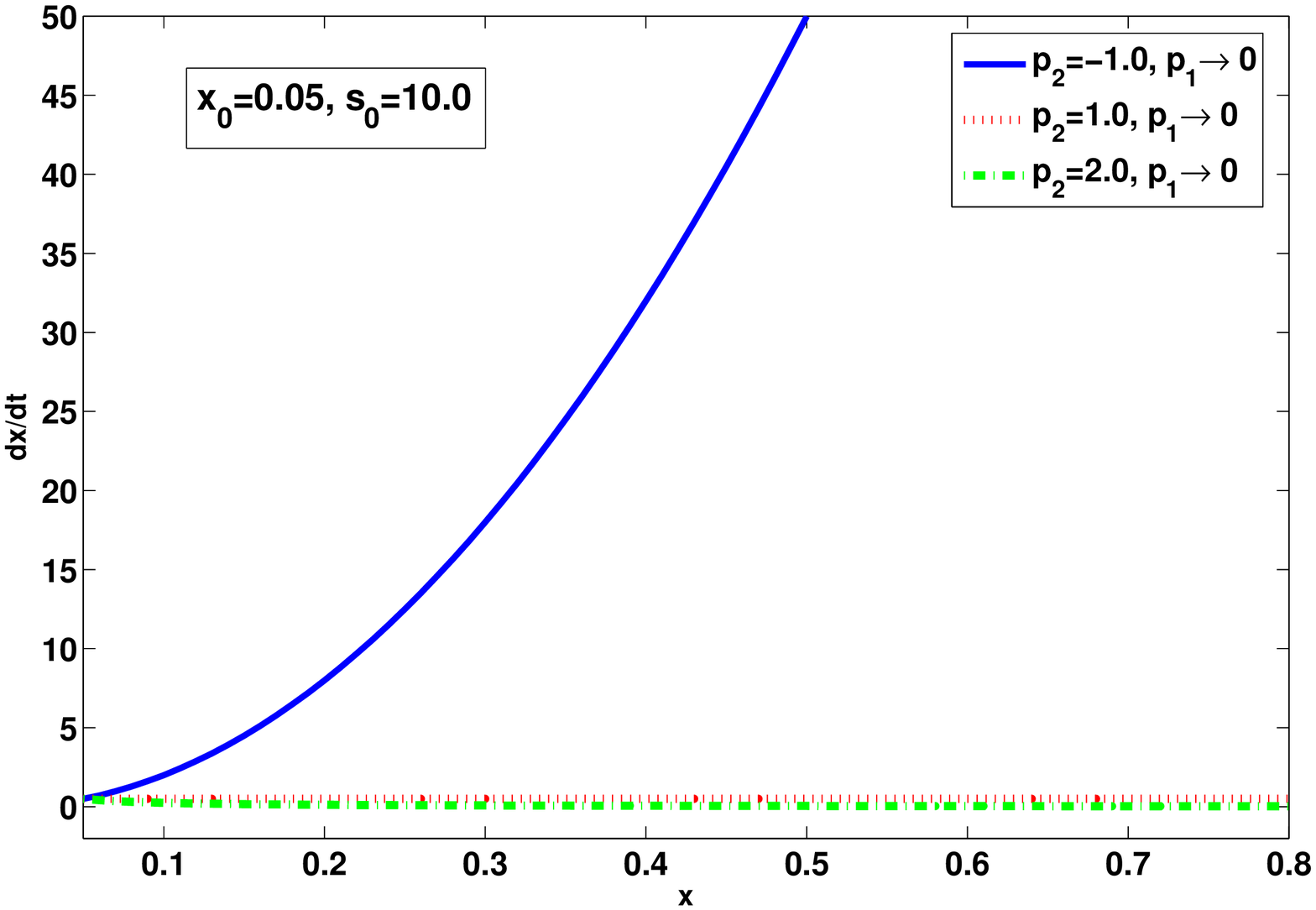}\\
  \caption{(Colour online) Plots of growth rate with respect to the state variable $x$ with $x_0=0.05$ and $s_0=10$ for $R=p_1s+p_2s^2$. From the top, the values of the parameter $p_2$ are $-1.0$ (potential growth for $p_1\rightarrow0$), $1.0$ (linear growth for $p_1\rightarrow0$) and $2.0$ (potential growth for $p_1\rightarrow0$) respectively. }
\label{figh:1}
\end{figure}
\begin{figure}
  \centering
  \includegraphics[width=4in, height=2in]{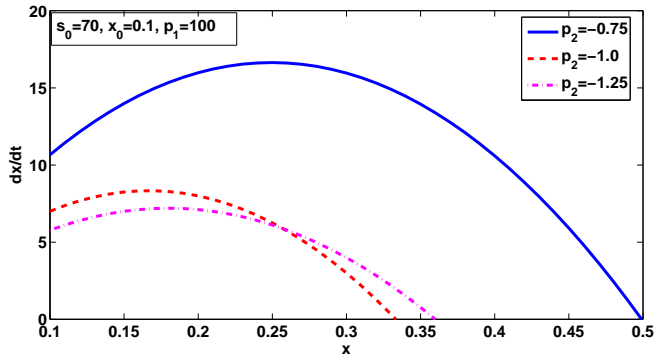}\\
  \caption{(Colour online) Plots of growth rate with respect to the state variable $x$ with $x_0=0.1$, $s_0=70$ and $p_1=100$ for $R=p_1s+p_2s^2$. From the top, the values of the parameter $p_2$ are $-0.75$, $-1.0$ and $-1.25$ respectively. Second characteristic curve from the top represents usual logistic growth. Others are $\theta-$ logistic growth by nature. $p_2$ indicates degree of competitive environment. The characteristic curve for $p_2=-1.25$ shows an increase in the saturation level of the state variable $x$ and decrease in the growth rate in a higher degree of competitive environment.}
\label{figh:2}
\end{figure}
\begin{figure}
  \centering
  \includegraphics[width=4in, height=2in]{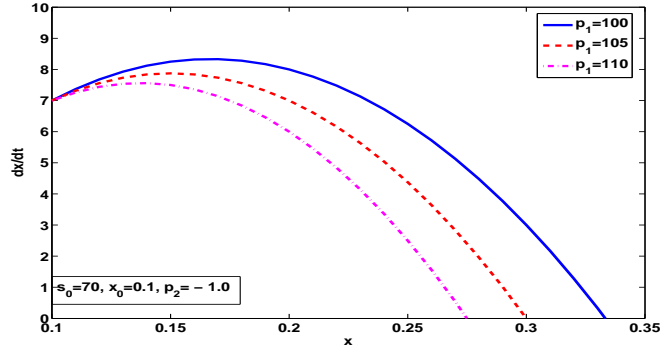}\\
  \caption{(Colour online) Plots of growth rate with respect to the state variable $x$ with $x_0=0.1$, $s_0=70$ and $p_2=-1$ for $R=p_1s+p_2s^2$. From the top, the values of the parameter $p_1$ are $100$, $105$ and $110$ respectively. All of them represent usual logistic growth by nature. An increase in $p_1$ lowers the optimum level, the saturation level and the growth rate of state variable $x$.}
\label{figh:3}
\end{figure}
\begin{figure}
  \centering
  \includegraphics[width=4in, height=2in]{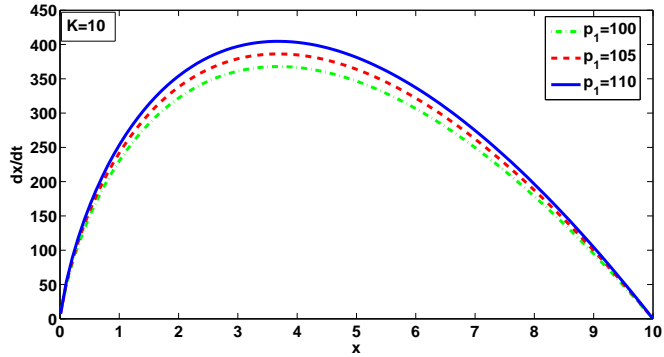}\\
  \caption{(Colour online) Plots of growth rate with respect to the state variable $x$ with $K=10$ and $p_2\rightarrow0$ for $R=p_1s+p_2s^2$. From the top, the values of the parameter $p_1$ are $110$, $105$ and $100$ respectively. All of them represent Gompertzian growth by nature. Optimum level and saturation level of state variable $x$ are independent of $p_1$. Optimum growth rate increases with the increase in $p_1$.}
\label{figh:4}
\end{figure}
\begin{figure}
  \centering
  \includegraphics[width=4in, height=2in]{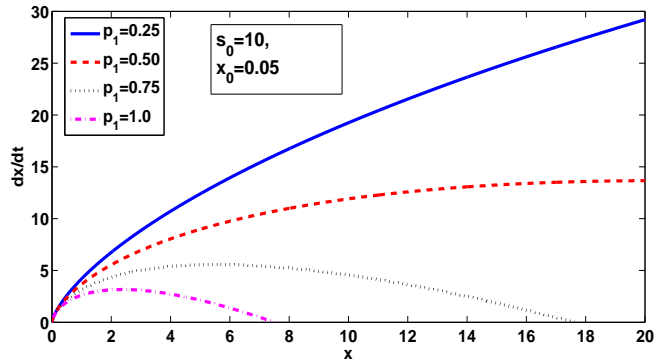}\\
  \caption{(Colour online) Plots of growth rate with respect to the the state variable $x$ with $x_0=0.05$, $s_0=10$ and $p_2=0.25$ for $R=p_1s+p_2s^2$. From the top, the values of the parameter $p_1$ are $0.25$, $0.5$, $0.75$ and $1.0$ respectively. All of them represent West-type biological growth processes. An increase in $p_1$ lowers the optimum level, the saturation level and the growth rate of state variable $x$.}
\label{figh:5}
\end{figure}
\begin{figure}
  \centering
  \includegraphics[width=4in, height=2in]{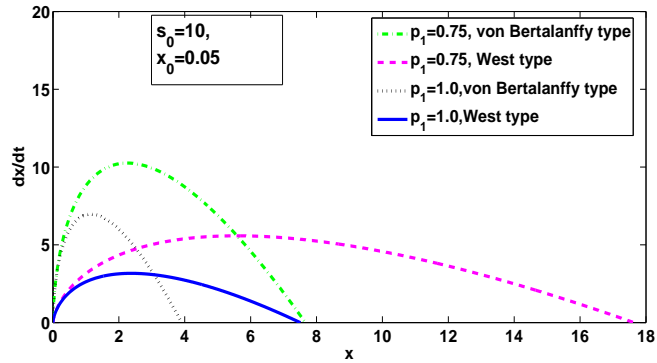}\\
  \caption{(Colour online) A comparative representation of von Bertalanffy type and West type growth for the same $p_1$ in terms of plots of growth rate with respect to the state variable $x$ with $x_0=0.05$, $s_0=10$ for $R=p_1s+p_2s^2$. First and third from the top represent von Bertalanffy type growth ($p_2$=0.34) for $p_1=0.75$ and $p_1=1.0$ respectively. Second and fourth characteristics curves from the top represent West type growth ($p_2$=0.25) for the same $p_1$. The optimum and saturation level of state variable $x$ of von Bertalanffy type growth is less than that of West type growth for same $p_1$, whereas growth rate of von Bertalanffy type growth is greater than that of West type growth for same $p_1$.}
\label{figh:6}
\end{figure}
\end{document}